\documentclass{appolb}
\usepackage{graphicx}
\usepackage{slashed}
\begin{document}
\title{Early quark deconfinement in compact star astrophysics and heavy-ion collisions %
\thanks{Presented at the $29^{\rm th}$ Conference ''Quark Matter 2022'' on ultrarelativistic nucleus-nucleus collisions, April 4-10, 2022, Krak\'ow, Poland}%
}
\author{
O. Ivanytskyi$^1$, D. Blaschke$^1$, T. Fischer$^1$, A. Bauswein$^2$
\address{$^1$Institute of Theoretical Physics, University of Wroclaw, 
Poland
\\
$^2$GSI Helmholtzzentrum f\"ur Schwerionenforschung, 
Darmstadt, Germany}
}
\maketitle
\begin{abstract}
Based on a recently developed relativistic density functional approach to color-superconducting quark matter and a novel quark-hadron transition construction which phenomenologically accounts for the effects of inhomogeneous pasta phases and quark-hadron continuity, we construct a class of hybrid equations of state applicable at the regimes typical for compact star astrophysics and heavy ion collisions. We outline that early quark deconfinement is a notable consequence of strong diquark pairing providing a good agreement with the observational data and driving the trajectories of the matter evolution during the supernovae explosions toward the regimes typical for the compact star mergers and heavy-ion collisions. 
\end{abstract}
  
\section{Introduction}

The puzzling question for the origin of compact stars (CS) with masses exceeding $2~{\rm M}_\odot$ can be successfully addressed at present only within the supernovae (SN) explosion mechanism based on quark deconfinement 
in the stellar matter \cite{Fischer:2017lag}. This serves as an indirect argument in favor of the existence of quark matter in cores of heavy CS. Binary CS mergers could produce a distinct postmerger gravitational wave signal \cite{Bauswein:2018bma}.
These interesting applications are summarised in \cite{Bauswein:2022vtq}.
They are based on a hybrid equation of state (EoS) that has been constructed from hadronic and quark matter EoS developed within relativistic density functional (RDF) approaches \cite{Kaltenborn:2017hus,Ivanytskyi:2022oxv}. In this contribution, we summarize recent developments of the RDF approach to quark matter which address beyond confinement also the aspects of chiral symmetry breaking and color superconductivity. In particular the occurrence of a large diquark pairing gap modifies the phase structure and EoS of QCD at low temperatures and is thus of central interest for the discussion of the existence and location of one or more critical endpoints (CEPs). A developed  constructive scheme generates thermodynamically consistent EoS with multiple or absent CEP and provides a solid basis for discussing their effects in simulations of astrophysical phenomena and heavy-ion collisions (HIC).

\section{Relativistic density functional for quark matter}

\begin{figure}[t]
\includegraphics[width=0.32\columnwidth]{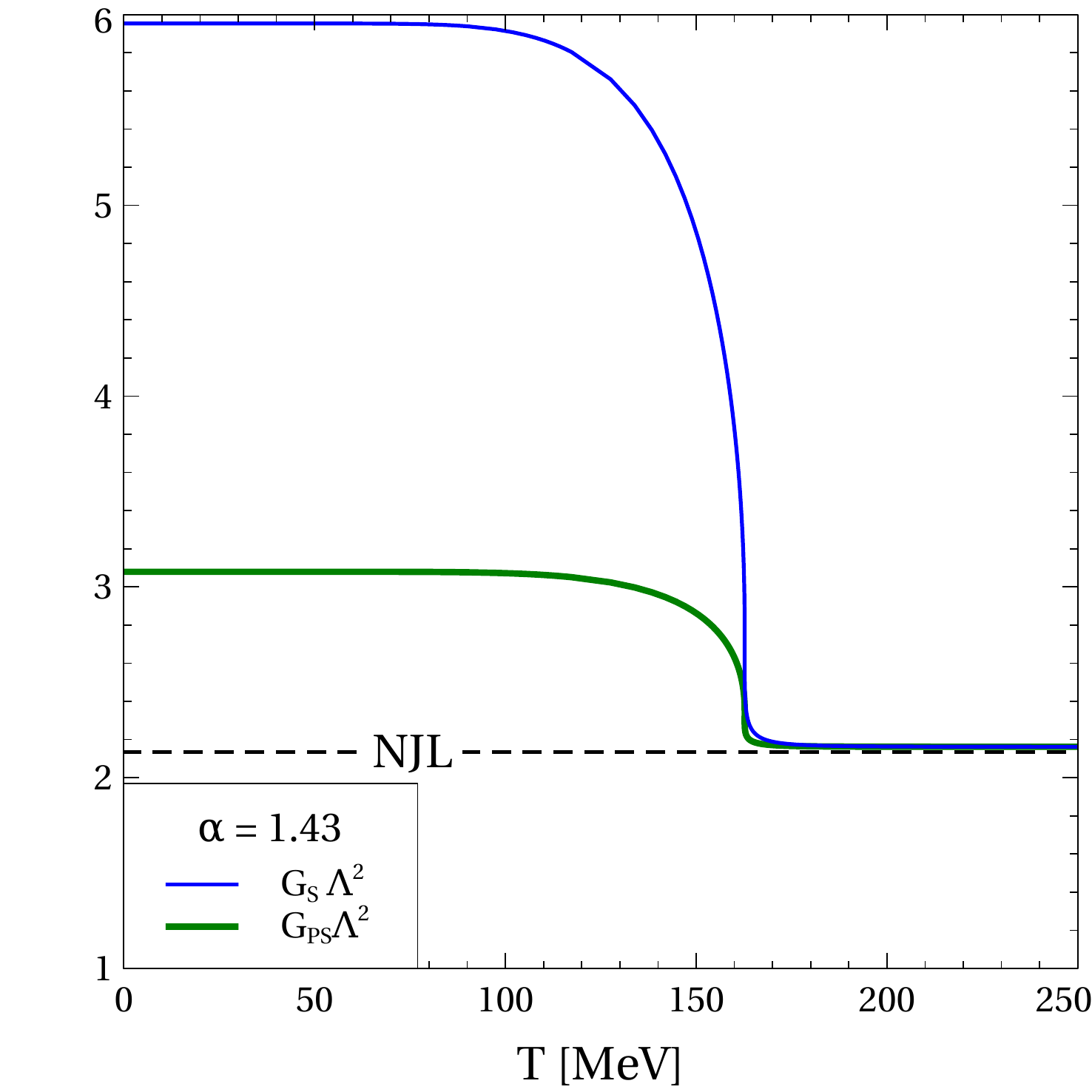}
\includegraphics[width=0.32\columnwidth]{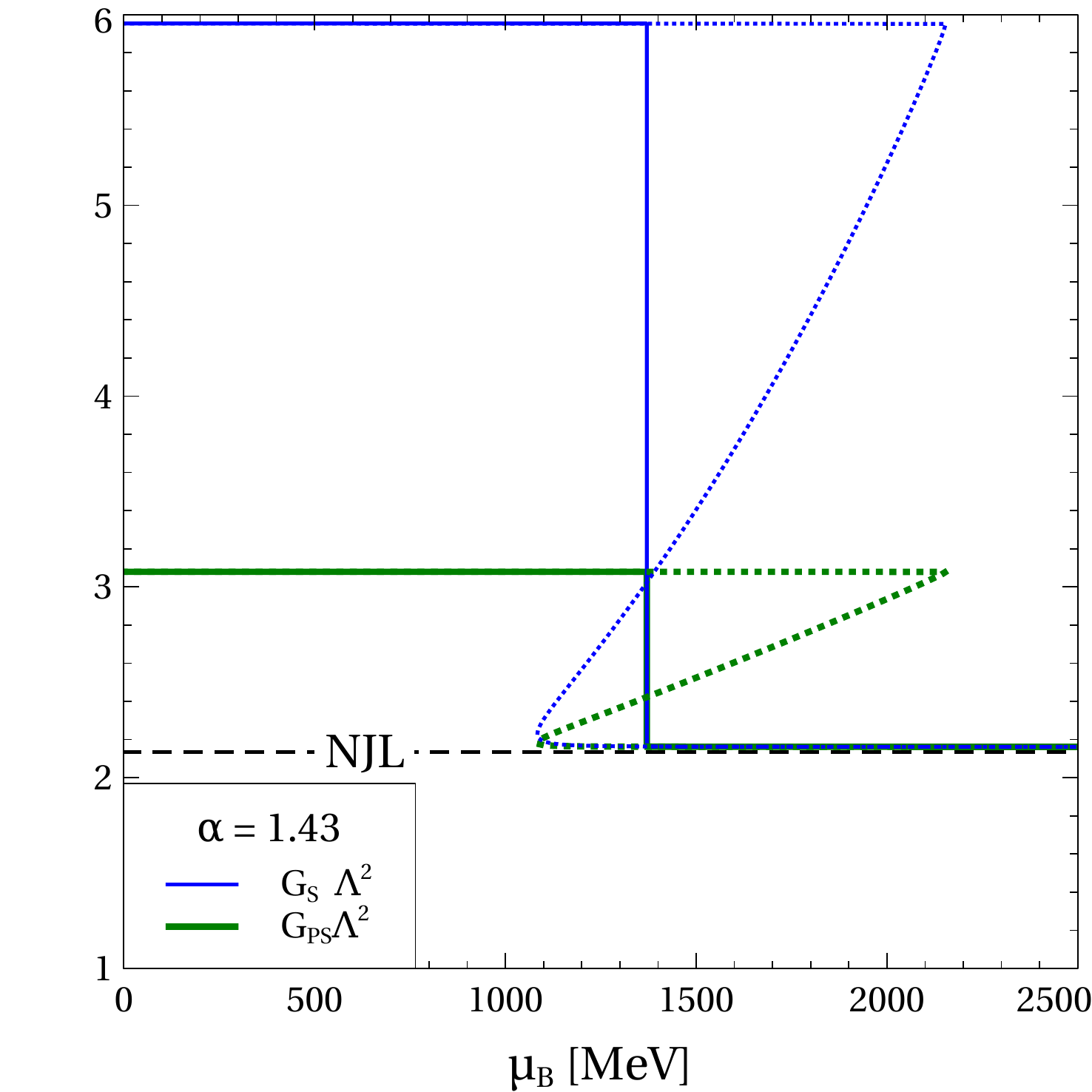}
\includegraphics[width=0.32\columnwidth]{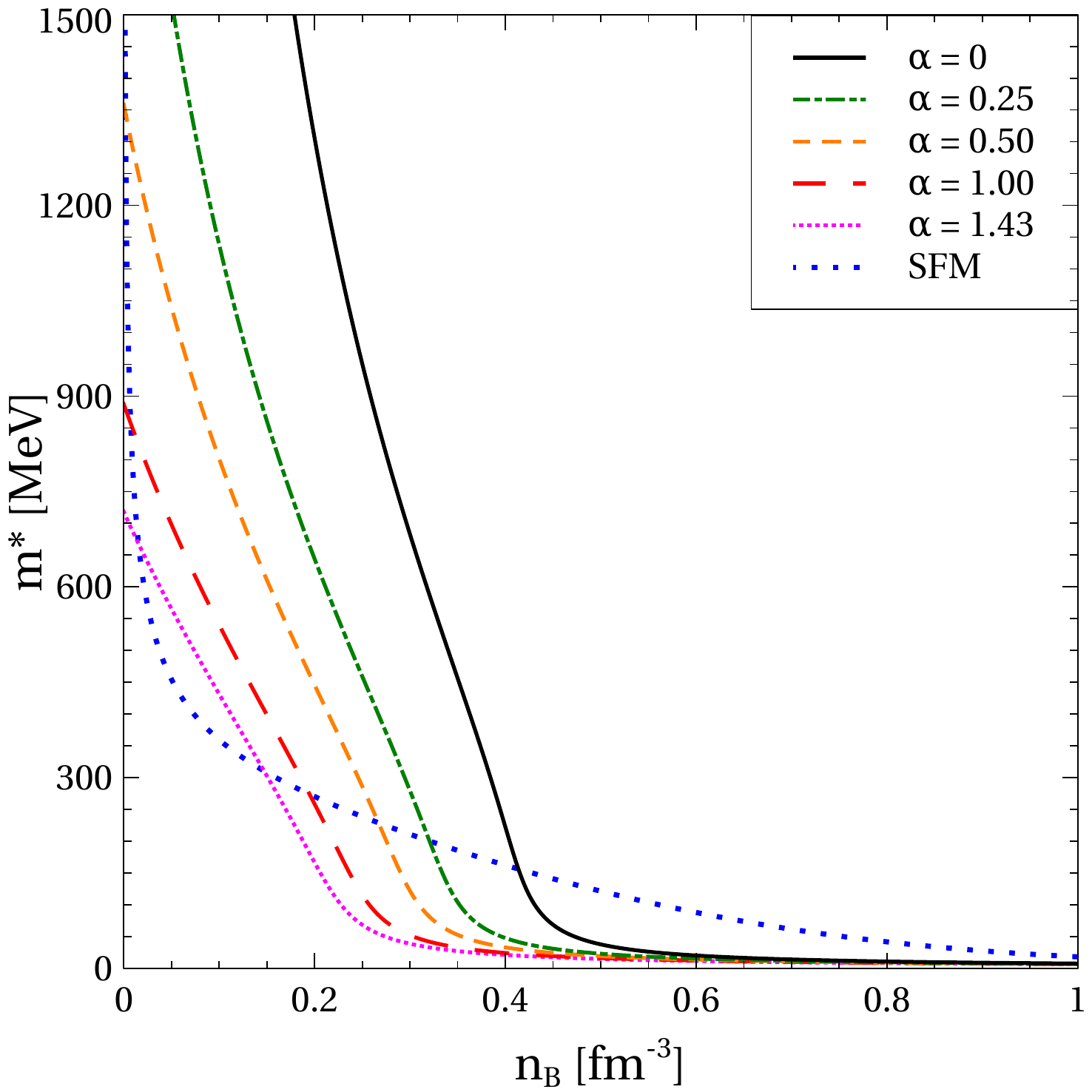}
\caption{Scaled effective scalar $G_S\Lambda^2$ and pseudoscalar $G_{PS}\Lambda^2$ couplings as functions of temperature $T$ at $\mu_B=0$ (left panel), baryonic chemical potential $\mu_B$ at $T=0$ (middle panel) and effective quark mass $m^*$ as function of baryon density $n_B$ (right panel) from Ref. \cite{Ivanytskyi:2022oxv}. Dashed lines on the left and middle panels represent the NJL value $G\Lambda^2=2.14$  \cite{Ratti:2005jh}. Dotted curves on the middle panel indicate the unstable parts that are removed by applying the Maxwell construction. The blue dotted line on the right panel is obtained within the SFM with $\alpha_{SFM}=0.39~{\rm fm}^{-3}$ \cite{Kaltenborn:2017hus}. Calculations are performed for symmetric quark matter, $G_V=G_D=0$, $\alpha$ specified in the legend and the rest of the model parameters with the values mentioned above.}
\label{fig1}
\end{figure}

The RDF approach from Ref. \cite{Ivanytskyi:2022oxv} is represented by the Lagrangian
\begin{eqnarray}
\label{I}
\mathcal{L}=\overline{q}(i\slashed\partial- m)q-
G_V(\overline{q}\gamma_\mu q)^2+
G_D(\overline{q}i\gamma_5\tau_2\lambda_A q^c)(\overline{q}^ci\gamma_5\tau_2\lambda_A q)-\mathcal{U}
\end{eqnarray}
with two-flavor quark field $q^T=(u~d)$, current quark mass $m$ and $G_V$, $G_S$ being coupling constants in vector repulsion and diquark pairing channels, respectively. A chirally symmetric generalization of the potential energy density functional inspired by the string-flip model (SFM) \cite{Kaltenborn:2017hus} reads
\begin{eqnarray}
\mathcal{U}&=&D_0\left[(1+\alpha)\langle \overline{q}q\rangle_0^2
-(\overline{q}q)^2-(\overline{q}i\gamma_5\vec\tau q)^2\right]^{\frac{1}{3}}\nonumber\\
\label{II}
&\simeq&\mathcal{U}_{MF}+
(\overline{q}q-\langle\overline{q}q\rangle)\Sigma_{MF}-
G_{S}(\overline{q}q-\langle\overline{q}q\rangle)^2-
G_{PS}(\overline{q}i\gamma_5\vec\tau q)^2.
\end{eqnarray}
Here $\alpha$ and $D_0$ are constants and $\langle \overline{q}q\rangle_0$ is the chiral condensate in the vacuum. The last line in Eq. (\ref{II}) corresponds to the second order expansion of $\mathcal{U}$ around the mean-filed solutions $\langle \overline{q}q\rangle$ and $\langle \overline{q}i\gamma_5\vec\tau q\rangle=0$ labeled with the subscript index ``$MF$''. This expansion brings the present model to the form of the NJL model with the mean-field scalar self-energy of quarks $\Sigma_{MF}=\partial\mathcal{U}_{MF}/\partial\langle\overline{q}q\rangle$ and effective couplings in scalar $G_S=-\partial\mathcal{U}_{MF}^2/\partial\langle\overline{q}q\rangle^2/2$ and pseudoscalar $G_S=-\partial\mathcal{U}_{MF}^2/\partial\langle\overline{q}i\gamma_5\vec\tau q\rangle^2/6$ channels. In Ref. \cite{Ivanytskyi:2022oxv} model parameters $m=4.2$ MeV, $\Lambda=573$ MeV, $\alpha=1.43$ and $D_0\Lambda^{-2}=1.39$ were fixed in order to reproduce the pion mass $M_\pi=140$ MeV and decay constant $F_\pi=92$ MeV, with the scalar meson mass $M_\sigma=980$ MeV and the vacuum value of the chiral condensate per flavor $\langle\overline{l}l\rangle_0=-(267~{\rm MeV})^3$. 
We note that $\Lambda$ is a three-momentum scale which occurs in the   smooth momentum cut-off by a Gaussian formfactor which regularizes divergent zero-point terms. The behavior of $G_S$ and $G_{PS}$ as well as the effective quark mass $m^*=m+\Sigma_{MF}$ is shown in Fig. \ref{fig1}. The dynamical breaking of chiral symmetry leads to $G_S\neq G_{PS}$ in the vacuum, while its dynamical restoration at high temperatures and/or densities is manifested by the asymptotic coincidence of the scalar and pseudoscalar couplings. This is reflected in the melt-down of $m^*$. Its vacuum value $m_0^*$ is controlled by the parameter $\alpha$ so, that $m_0^*\rightarrow\infty$ at $\alpha\rightarrow0$. For the mentioned set of parameters $m_0^*=718$ MeV and the pseudocritical temperature at $\mu_B=0$ defined by the peak of the chiral susceptibility is 163 MeV. 
The quark matter EoS is obtained by treating the present model within the mean-field approximation.
It is remarkable that the BCS relation between the mass gap in the vacuum and the critical temperature for its restoration, which holds for the (P)NJL model in the chiral limit, is violated for this class of quark matter models. 

\section{Phase diagram of strongly interacting matter}

High values of the effective quark mass at low $T$ and $\mu_B$ represent phenomenological confinement in the RDF approach. This makes description of strongly interacting matter in terms of quark degrees of freedom inadequate in the confinement region and requires matching the quark matter EoS to the hadron one yielding a hybrid quark-hadron EoS. Within the Maxwell construction of quark-hadron transition the matching point is defined by the baryon chemical potential $\mu_B^{\rm Max}$ at which the pressures of two phases coincide, while the baryon density discontinuously jumps from $n_B^h|_{\rm Max}$ on the hadron side to $n_B^q|_{\rm Max}$ on the quark one. This picture ignores inhomogeneous structures in the quark-hadron interface known as pasta phases \cite{Maslov:2018ghi} and corresponds to a sharp interface between two phases. Accounting for those pasta phases would wash out the sharp quark hadron interface allowing for the existence of a mixed phase, which  is restricted by the baryon chemical potentials $\mu_B^h$ and $\mu_B^q$ (corresponding to $n_B^h$ and $n_B^q$) from the hadron and quark sides, respectively. 
In Ref. \cite{Ayriyan:2021prr}, the EoS of the mixed phase was parameterized by two pieces of parabolic functions. In Ref. \cite{Ivanytskyi:2022wln} such a two-zone interpolation scheme (TZIS) was further developed to the case of arbitrary fractions of electric charge and applied at finite temperatures. 
The parameters of these two parabolic functions were defined so that both the pressure $p$ and the baryon density $n_B$ remain continuous at the mixed phase boundaries. Continuity of $p$ is also required at the matching point of two parabolas $\mu_B^c=(\mu_B^h+\mu_B^q)/2$, while $n_B$ experiences a discontinuous jump of $\Delta n_B$. The TZIS is given a closed form with the parameterization 
\begin{eqnarray}
\label{III}
&&\mu_B^h=\mu_B^{\rm Max}|_{T=0}(1-x)\sqrt{1-T^2/T_0^2},\quad
\mu_B^{q}=\mu_B^{\rm Max}(1+x),\\
\label{IV}
&&\Delta n_B=n^*(T_{cep1}-T)^\beta(T-T_{cep2})^\beta\theta(T_{cep1}-T)\theta(T-T_{cep2}),
\end{eqnarray}
\begin{figure}[t]
\centering
\includegraphics[width=0.32\columnwidth]{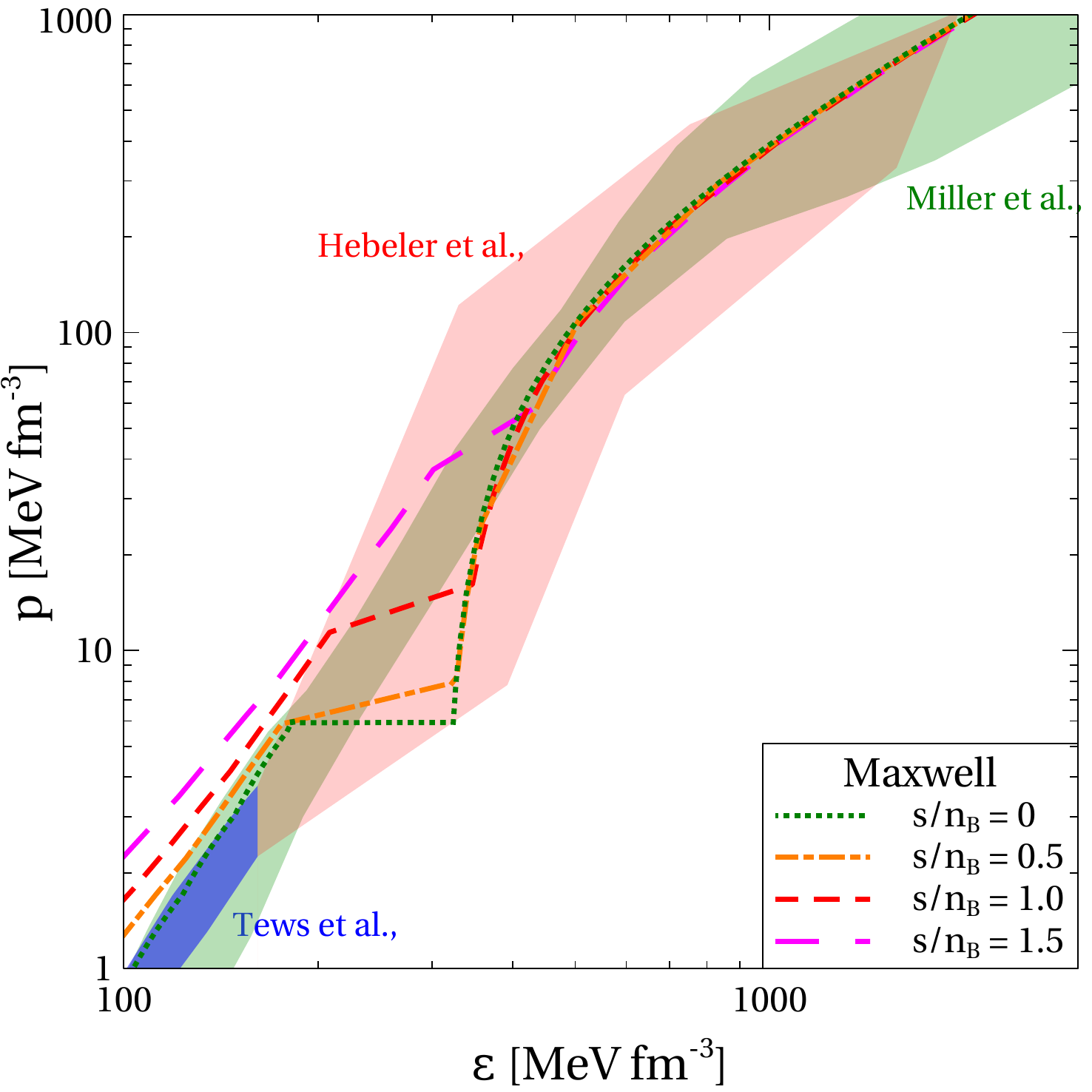}
\includegraphics[width=0.32\columnwidth]{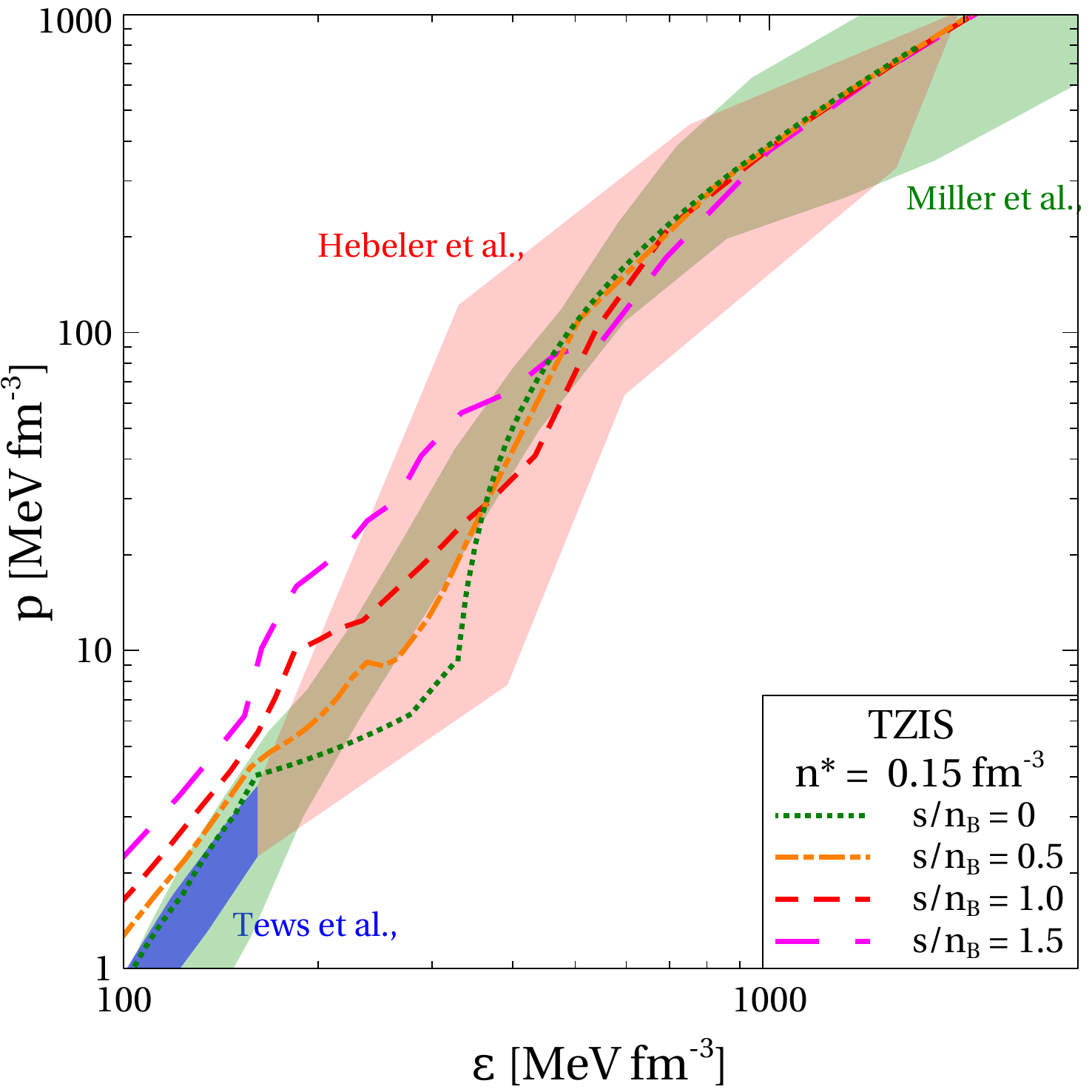}
\includegraphics[width=0.32\columnwidth]{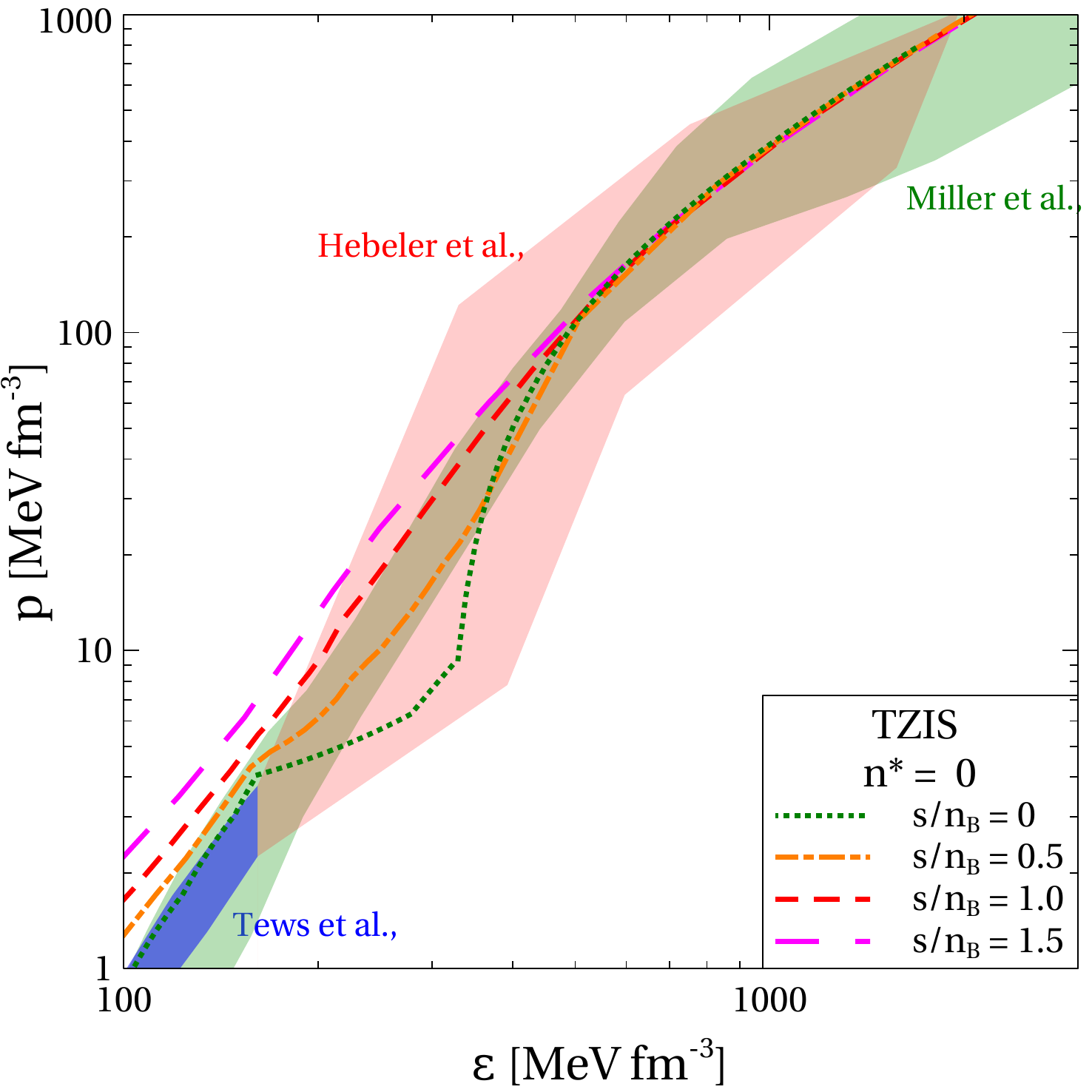}
\caption{Pressure $p$ of electrically neutral $\beta$-equilibrated quark-hadron matter vs. energy density $\varepsilon$ along the isentropes $s/n_B=const$ found within the Maxwell construction (left panel) and the TZIS with $n^*=0.15~{\rm fm}^{-3}$ (middle panel) and $n^*=0$ (right panel). The shaded areas represent the cold nuclear matter constraints.}
\label{fig2}
\end{figure}
where $x=0.01$, $n^*=0$ or 0.15 fm$^{-3}$, $T_{cep1}=90$ MeV and $T_{cep2}=15$ MeV correspond to high and low temperature CEPs and $\beta=0.3265$ is the critical exponent of the 3D Universality class \cite{Campostrini:2002cf}. The TZIS allows us to construct a hybrid quark-hadron EoS at arbitrary entropy per baryon $s/n_B$. Fig. \ref{fig2} compares such an EoS to the one obtained within the Maxwell construction. Furthermore, having the edges of the mixed quark-hadron phase defined we can construct the phase diagram of strongly interacting matter that is shown in Fig. \ref{fig3}. It is remarkable that the transition from quark to hadron matter leads to a growth of $T$ along adiabates $s/n_B=const$ being a direct consequence of the reduction of the number of accessible microstates due to the transition to the color superconducting phase of quark matter \cite{Ivanytskyi:2022oxv}.

\begin{figure}[t]
\centering
\includegraphics[width=0.32\columnwidth]{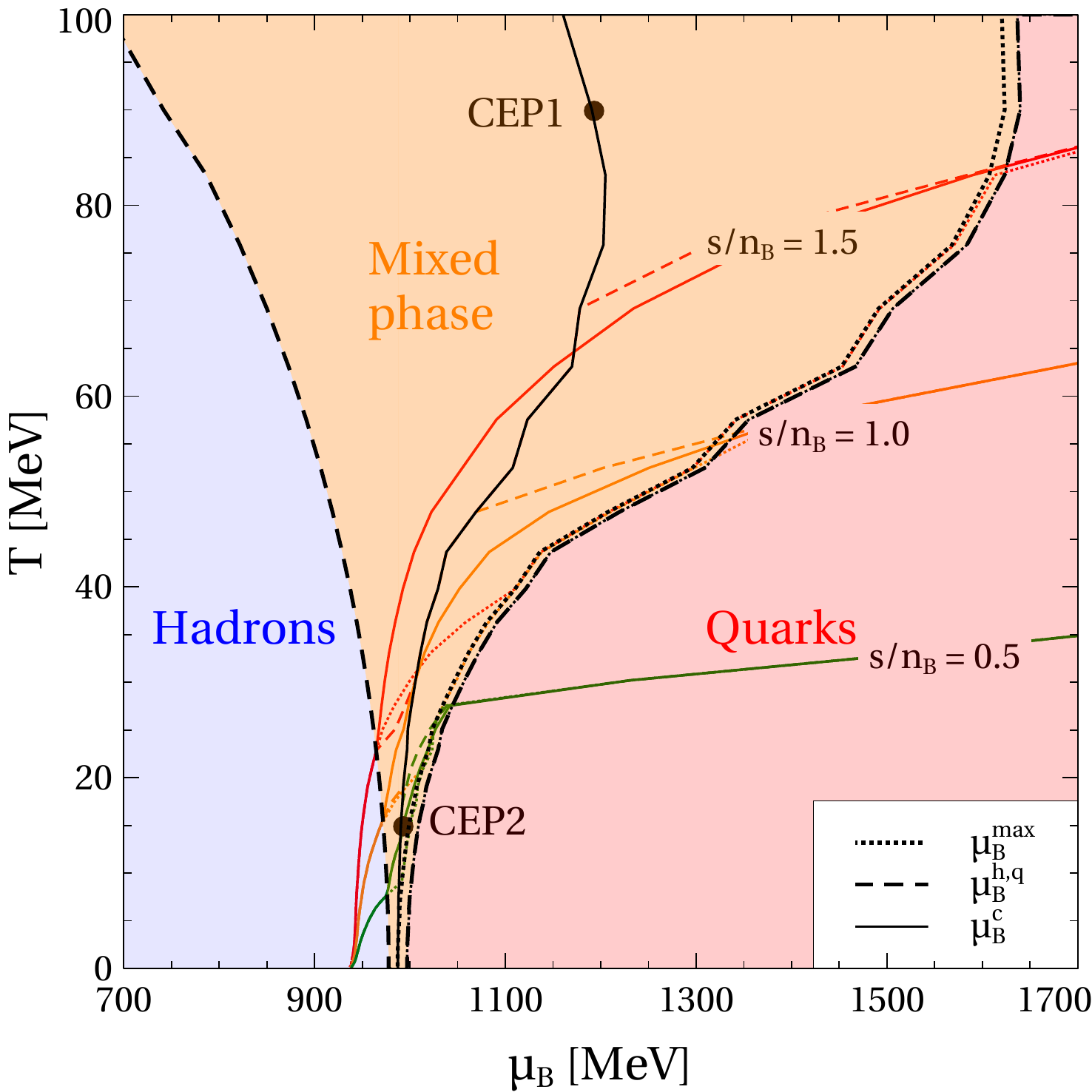}
\includegraphics[width=0.32\columnwidth]{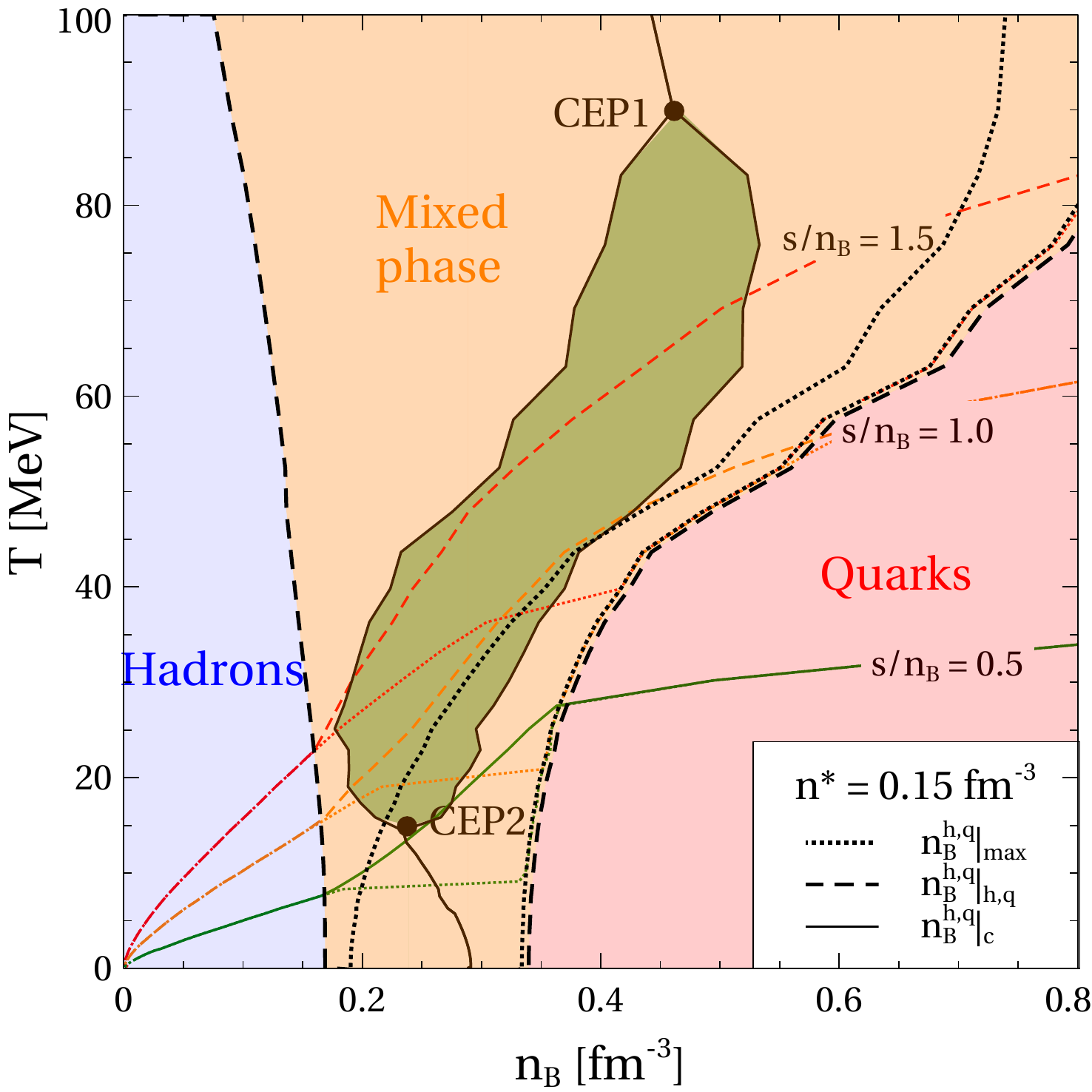}
\includegraphics[width=0.32\columnwidth]{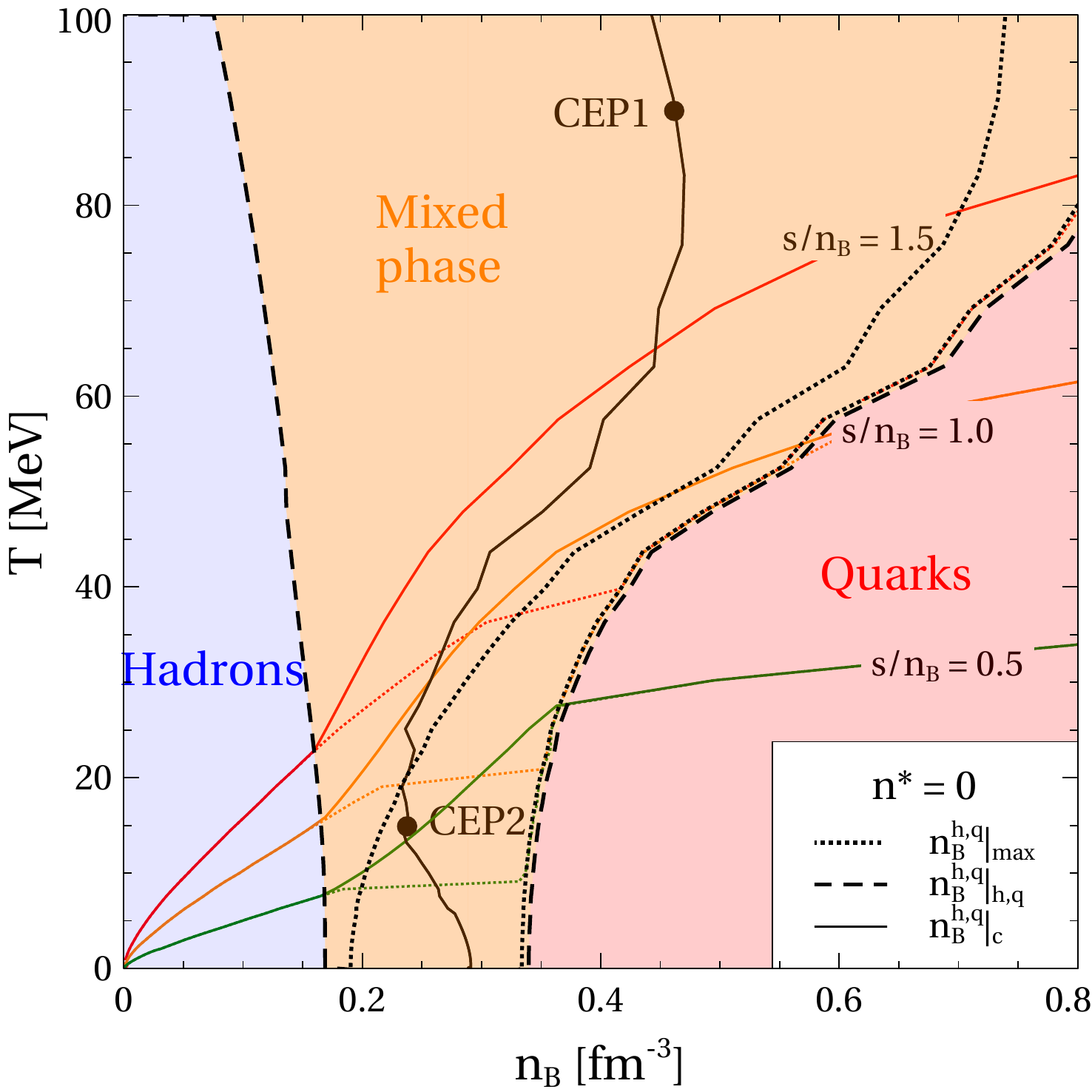}
\caption{Phase diagram of $\beta$-equilibrated electrically neutral quark-hadron matter in the $\mu_B-T$ (left panel) and $n_B-T$ (central and right panels) planes. 
The black dotted, dashed and solid curves correspond to the phase boundaries and the matching chemical potential discussed in the text. 
The color mapping of phases corresponds to the TZIS. The filled black circles show the CEPs, which on the right panel are to guide the eye. 
The colored solid, dashed and dotted curves show adiabates $s/n_{B}=const$ calculated within the Maxwell construction, the TZIS with $n^*=0.15~{\rm fm}^{-3}$ and $n^*=0$, respectively.
The green shaded area shows the region where $n_B$ discontinuously jumps within the TZIS with $n^*=0.15~{\rm fm}^{-3}$.}
\label{fig3}
\end{figure}

\section{Compact stars at vanishing and finite entropy}

\begin{figure}[t]
\centerline{%
\includegraphics[width=0.32\columnwidth]{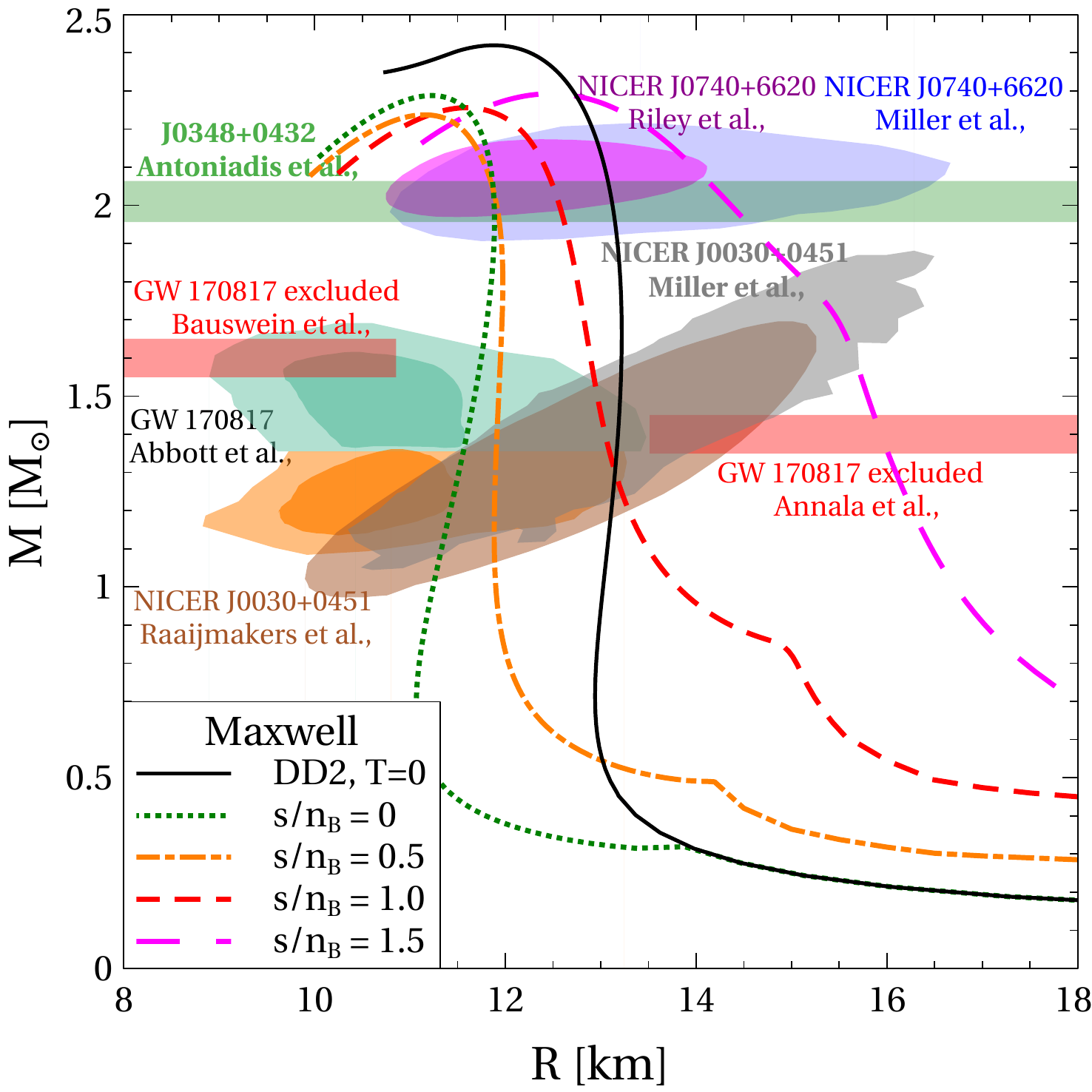}
\includegraphics[width=0.32\columnwidth]{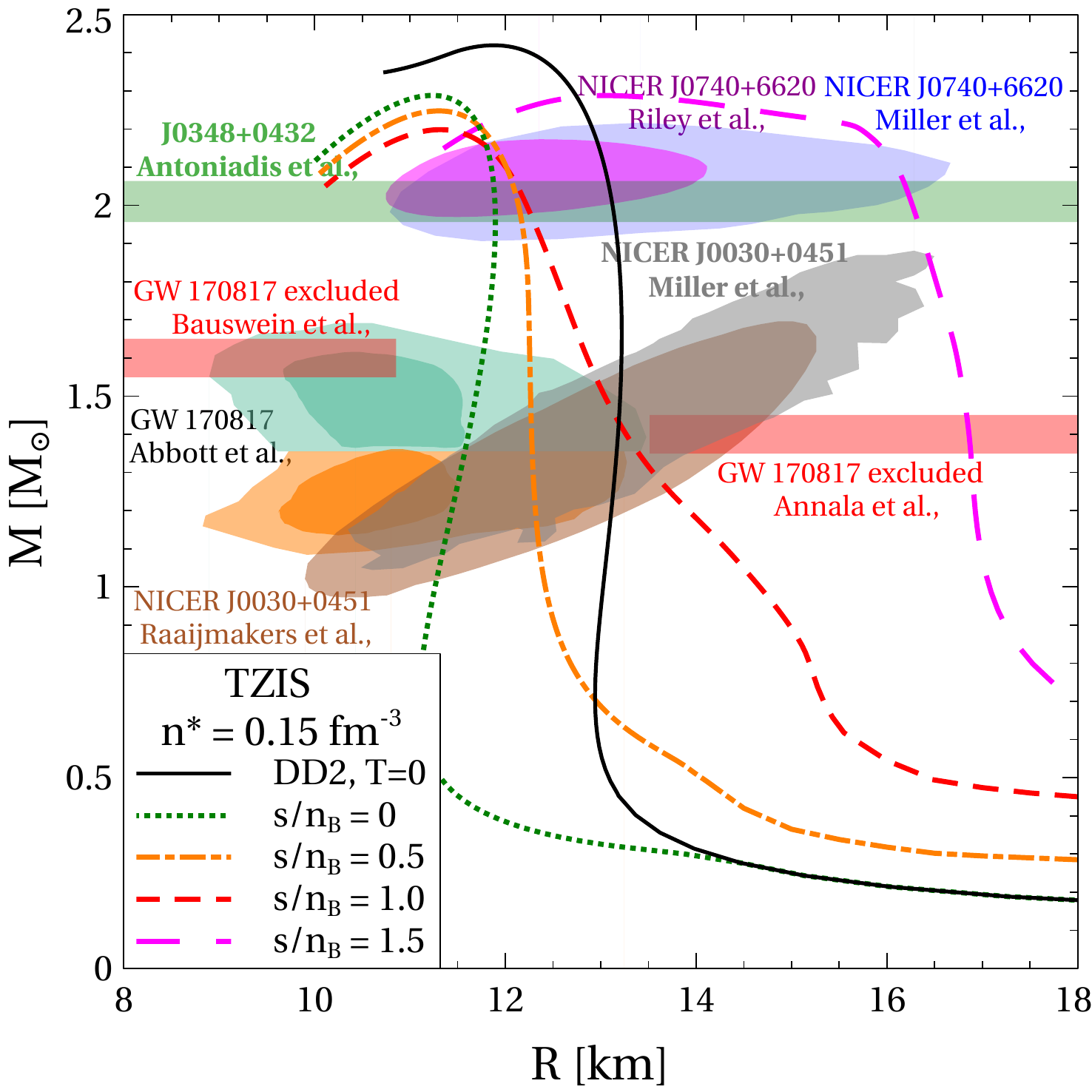}
\includegraphics[width=0.32\columnwidth]{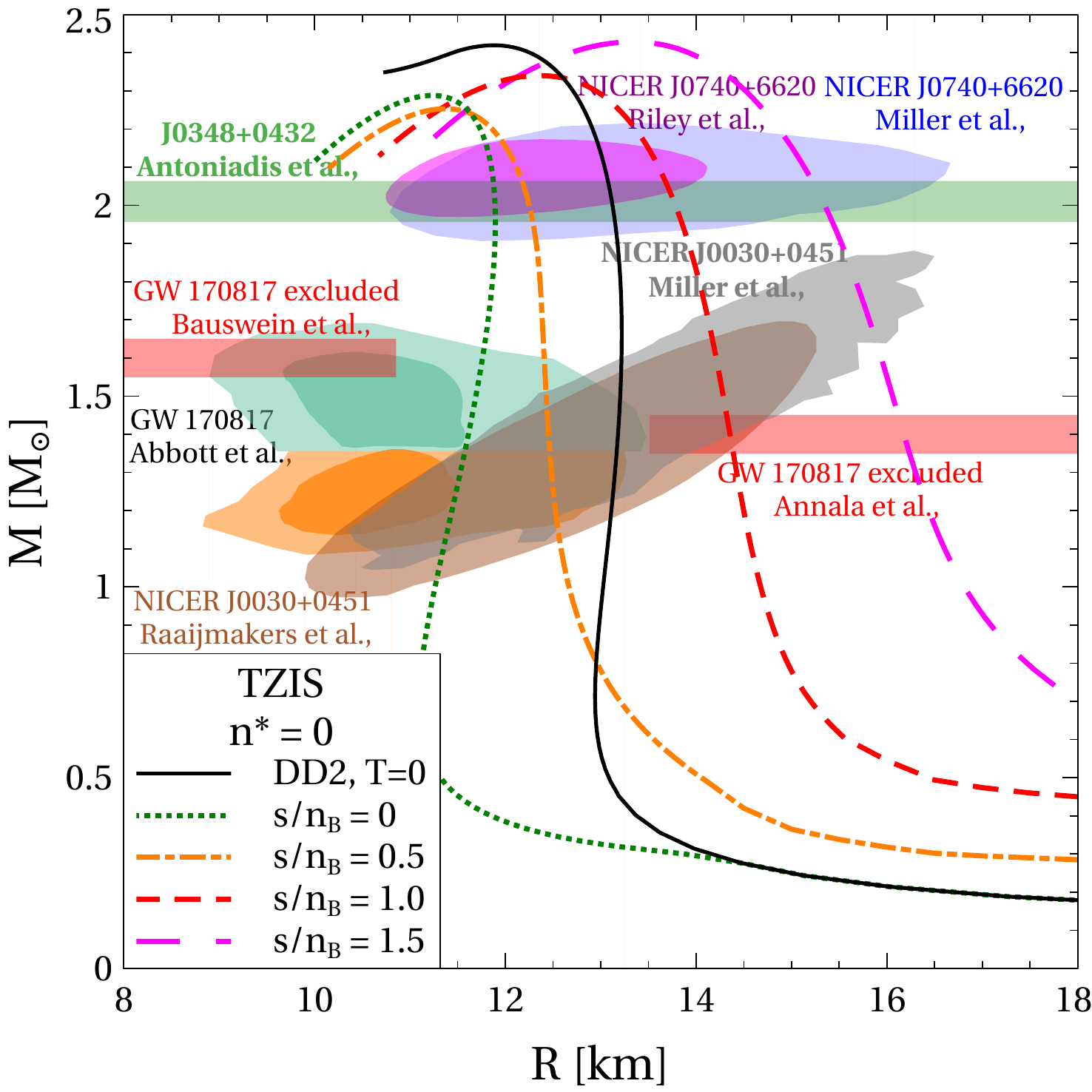}}
\caption{Mass-radius relation of hybrid NS with the isentropic quark-hadron EoS presented in Fig. \ref{fig2}. Black solid curves obtained with the DD2 EoS of cold hadron matter are given for the sake of comparison. The astrophysical constraints depicted by the colored bands and shaded areas correspond to the case of cold neutron stars.}
\vspace*{-.2cm}
\label{fig4}
\end{figure}

Entropy of the quark-hadron matter in the interiors of the proto NS remains approximately constant during SN explosions \cite{Fischer:2017lag}. Therefore, isentropic EoS of quark-hadron matter is phenomenologically interesting. We applied such EoSs shown in Fig. \ref{fig2} to solving a problem of relativistic hydrostatic equilibrium \cite{Ivanytskyi:2022wln}. The corresponding mass radius relations of cold NS ($s/n_B=0$) and warm proto-NS ($s/n_B\neq0$) are shown in Fig. \ref{fig4}. In the case of cold NS our approach provides agreement with the constraints from Refs. \cite{
Riley:2021pdl,Miller:2021qha,Riley:2019yda,
LIGOScientific:2018cki,Bauswein:2017vtn,Annala:2017llu} and gives the tidal polarizability of $1.4~{\rm M}_\odot$ mass stars $\Lambda_{1.4}=540-550$ agreeing with Ref. \cite{LIGOScientific:2018cki}. Finite $s/n_B$ increases the radius of NS but leaves their maximal mass almost unchanged.

\section{Conclusions}

We developed a confining RDF for color-superconducting quark matter and produced a family of hybrid quark-hadron EoS with or without (multiple) CEP(s). Due to large values of the diquark pairing gap our approach favors early quark deconfinement, provides good agreement with the present astrophysical constraints and drives trajectories of the evolution of stellar matter during the SN explosions toward the temperatures range of HIC.

\subsection*{Acknowledgements}
This work was supported by 
NCN under grants 2019/33/B/ST9/03059 (O.I., D.B.) and 2020/37/B/ST9/00691 (T.F.).
A.B. acknowledges support by the European Research Council 
under the European Union's Horizon 2020 research and innovation program, grant No. 759253, by 
DFG Project-ID 279384907 - SFB 1245, by DFG - Project-ID 138713538 - SFB 881 
and by the State of Hesse within the Cluster Project ELEMENTS.
The work was performed within a project that has received funding from the Horizon 2020 program under grant agreement STRONG-2020 - No. 824093.


\bibliography{refs}

\end{document}